\documentclass[12pt,journal,onecolumn,draftcls]{IEEEtran}

\newcommand{\textred}[1]{\textcolor{red}{#1}}

\ifx\noeditingmarks\undefined
  \newcommand{\pgwrapper}[2]{\textred{#1: #2}}
\else
  \newcommand{\pgwrapper}[2]{}
\fi
\usepackage{soul}
\usepackage{framed}
\usepackage{bbm}
\usepackage{epsfig}
\usepackage{times}
\usepackage{float}
\usepackage{afterpage}
\usepackage{amsmath}
\usepackage{amstext}
\usepackage{amssymb,bm}
\usepackage{latexsym}
\usepackage{color}
\usepackage{graphicx}
\usepackage{amsmath}
\usepackage{amsthm}
\usepackage{enumitem}
\usepackage{graphicx}
\usepackage{subfigure}
\usepackage[center]{caption}
\usepackage{pstricks}
\usepackage{caption}
\usepackage{booktabs}
\usepackage{multicol}
\usepackage{lipsum}
\usepackage{verbatim}
\usepackage{hyperref}
\usepackage[normalem]{ulem}

\newtheorem{remark}{Remark}

\newtheorem{lemma}{Lemma}
\newtheorem{definition}{Definition}
\newtheorem{theorem}{Theorem}
\newtheorem{example}{Exmple}

\newcommand{\hq}{\widehat{Q}}
\newcommand{\prob}{\mathbbm{P}}
\newcommand{\PP}{\mathbb{P}}

\newcommand{\Qle}{Q_{\lambda+\epsilon}}

\begin{document}

\title{On the Capacity of the Slotted Strongly Asynchronous Channel with a Bursty User
}
\author{
\IEEEauthorblockN{
Sara Shahi, Daniela Tuninetti and Natasha Devroye\\}
\IEEEauthorblockA{%
University of Illinois at Chicago, Chicago IL 60607, USA.
Email: {\tt sshahi7, danielat, devroye @uic.edu}}%
}

\maketitle

\begin{abstract}
In this paper, the trade-off between the number of transmissions (or  burstiness) $K_n=e^{n\nu}$ of a user, the asynchronism level $A_n=e^{n\alpha}$ in a slotted strongly asynchronous channel, and the ability to distinguish $M_n=e^{nR}$ messages per transmission with vanishingly error probability is investigated in the asymptotic regime as blocklength $n$ goes to infinity.
The receiver must locate and decode, with vanishing error probability in $n$, all of the transmitted messages.  
Achievability and converse bounds on the trade-off among $(R,\alpha,\nu)$ is derived. For cases where $\nu=0$ and $ R=0$, achievability and converse bounds coincide. A second model for a bursty user with random access in which the user may access and transmit a message in each block with probability $e^{-n\beta}$ in then considered. Achievability and converse bounds on the trade-off between $(R, \alpha, \beta)$ is also characterized. For cases where $\beta =\alpha$ and $R=0$, the achievability and converse bounds match.
\end{abstract}

\section{Introduction}
It is widely believed that Machine-type Communications and Internet of Things are going to be the next dominant paradigm in wireless technology. The traffic pattern imposed by the devices within these  networks  have unique features different from the ones in human-type communication networks.
The communications that take place within these networks are often sporadic and bursty, but must nonetheless be reliably detected and decoded. For example, each sensor node may want to transmit a signal to the base station only when some incident has taken place. 

In this paper, we consider the problem of both detecting and decoding asynchronous data bursts of a single user. This extends work in~\cite{shahi-itw2017}. In conventional methods the user transmits a pilot signal at the beginning of each data burst to notify the decoder of the upcoming data; the decoding phase may be performed using any synchronized decoding method. The loss in this approach is negligible when synchronization is done once and the cost of acquiring synchronization is  absorbed into the lengthy data stream that follows. For sparse / bursty transmission, as in the problem considered here, this approach is not suitable as the training based schemes are known  to be sub-optimal~\cite{aslan-suboptimality}.
In this work we do not enforce the usage of pilot symbols, and the codebook serves the dual purpose of synchronization and data transfer. 
 This paper's central goal is to characterize the trade-off between the reliable transmission rate between one transmitter and one receiver, the burstiness of that transmitter, and the level of asynchronism.

\subsection{ Past Work}
 In this paper, the trade-off between the number of transmissions (or  burstiness) $K_n=e^{n\nu}$ of a user, the asynchronism level $A_n=e^{n\alpha}$ in a slotted strongly asynchronous channel, and the ability to distinguish $M_n=e^{nR}$ messages per transmission with vanishingly error probability is investigated in the asymptotic regime as blocklength $n$ goes to infinity.
 The problem considered here  generalizes the one in~\cite{Strong_P2P-it}. In \cite{Strong_P2P-it}, the authors considered a user who transmits only once within an strong asynchronous window. The goal was to locate and decode the user transmission time and message. 
 In our work, the user transmits exponentially many times in blocklength (or arbitrary number of times in our second model). Moreover our error metric is the  global / joint probability of error (i.e., an error is declared if  {\it any} of the user's transmissions is in  error, and we have an exponential number of transmissions) and we require the exact recovery of the transmission time and codeword in {\it all} transmissions.  The approach in~\cite{Strong_P2P-it} does not extend to  the global probability of error criterion for exponential number of transmission in blocklength $n$ (where the number of transmissions is equal to $K_n=e^{n\nu}, \nu>0$). This is due to the fact that their achievability relies on the typicality decoder and the derived error bounds do not decay fast enough with blocklength $n$. 

In~\cite{optimal_sequential}, the authors considered the special case of the problem considered here where a user transmits one  synchronization pattern of length $n$ (hence the rate $R=0$) only once (hence $\nu=0$) in a window  of length $A_n=e^{n\alpha}$ of $n$ channel uses each. They showed that for any $\alpha$ below the synchronization threshold $\alpha_0$, the user can  detect the location of the synchronization pattern. In addition they showed that{ a  synchronization pattern consisting of the repetition of a single symbol which induces an output distribution with  the maximum divergence from the noise distribution, suffices.}
  The typicality decoder introduced in~\cite{optimal_sequential} however, even in a slotted channel model, only retrieves one of the trade-off points that we obtain in this paper that corresponds to a sub-exponential number of transmissions.  
  We propose   new achievability and converse techniques to support an exponential number of transmissions ($K_n=e^{n\nu}, \nu>0$). {Interestingly, we show that the symbol used for synchronization may change for different  values of $\alpha$ and $\nu$.}

The single user strongly asynchronous channel was also considered in~\cite{polyanskiy}, where it was shown that the exact transmission time recovery, as opposed to the error criterion in~\cite{Strong_P2P-it} which allows a sub-exponential delay in $n$, does not change the capacity.

Recently, the {\it synchronous} Gaussian massive multiple access channel with random access has been modeled in~\cite{guo-Many_Access} where the number of users is let to grow linearly in the code blocklength and any random subset of users may try to access the channel. In~\cite{guo-Many_Access}, the authors took advantage of the Gaussian channel structure to exactly derive matching upper and lower bounds on the capacity. Since then, other versions of ``massive number of users" have been proposed in~\cite{polyanskiy_ManyAccess},~\cite{liu2017massive}. In~\cite{shahi2016isit}, we studied  a multi-user version of the slotted strongly {asynchronous}  model  for a  discrete memoryless channel {where we assumed that $K_n=e^{n\nu}$ different users transmit  a message among $M_n^{(i)}=e^{nR_i}, i\in  \{1,\ldots,K_n\}$ of them only once in an asynchronous window of length $A_n=e^{n\alpha}$ blocks of $n$ channel uses each}. {Inner  bounds on the trade-off between $(R_i, \nu,\alpha), i\in \{1,\ldots,K_n\}$ were derived, but these were not shown to be tight.} 
 What renders the presented version of the problem {-- a single user transmitting multiple times rather than multiple users transmitting once each --} more tractable is that one is guaranteed that in each block there is at most one transmitted message and we do not need to detect the user's identity. 
 
\subsection{ Contributions} 
In this paper, we bridge the bursty random access channel model with the asynchronous communication and study the trade-off between the number of transmissions (or  burstiness) $K_n=e^{n\nu}$ of a user, the asynchronism level $A_n=e^{n\alpha}$ in a slotted bursty and strongly asynchronous channel, and the ability to distinguish $M_n=e^{nR}$ messages per transmission with vanishingly error probability as blocklength $n$ goes to infinity.
The slotted assumption restricts the transmission times to be integer multiples of the blocklength $n$; this assumption simplifies the error analysis yet captures the essence of the problem. %
 We show:
\begin{enumerate}
\item 
For synchronization and data transmission ($R> 0$), we find converse and achievability bounds on the capacity region of $(R, \alpha, \nu)$ that match for $\nu=0$.

\item
For synchronization only ($R=0$), our proposed sequential decoder achieves the optimal trade-off.  {Surprisingly, we show that the optimal synchronization pattern is not fixed and may depend on the asynchronism level $\alpha$.}

\item  For certain values of $R$, which are small enough, the achievability and converse bounds match. 

\end{enumerate}

We also consider a slotted bursty and strongly asynchronous random access channel with asynchronous level $A_n=e^{n\alpha}$ where the number of transmissions of the user is not fixed and the user may randomly with probability $p_n=e^{-n\beta}$ transmit a message, among $M_n=e^{nR}$ possible ones, within each block of $n$ channel uses. In this case, we show:
\begin{enumerate}[resume]
\item The achievability and converse bounds on the capacity region $(R, \alpha,\beta)$ is derived.  Our achievability result shows that the asynchronous window length $A_n=e^{n\alpha}$ will increases with the increase of $\beta$ since the number of transmissions to be detected decreases. Moreover, for $\beta = \alpha$ and $R=0$ the achievability and converse bounds match.
\end{enumerate} 
\subsection{Paper organization} The rest of this paper is organized as follows. 
 In Section~\ref{sec: sync+data} we introduce the slotted bursty and strongly asynchronous channel model with fixed number of transmissions and analyze its capacity region by providing upper and lower bounds on its capacity. We also find an equivalent capacity region expression for the special case with zero rate (synchronization only).  In Section~\ref{sec:random transmission} we introduce a model for slotted bursty and strongly asynchronous channel with random number of transmissions and find upper and lower bounds on its capacity region. Section~\ref{sec:conclusion} concludes the paper. 

\subsection{ Notation}\label{subsec:notation}
The notation $a_n\doteq e^{nb}$ means $\lim_{n\to \infty}\frac{\log a_n}{n}=b$. 
We write $[M:N]$, where $ M, N\in \mathbb{ Z}, M\leq N$, to denote the set $\{M, M+1, \ldots, N\}$, and $[K]:=[1:K]$. In addition we use the notation \[[a]^+:= \left\{\begin{matrix} a, & a>0\\ 0, & a\leq 0 \end{matrix}\right. .\]

Capital letters represent  random variables that take on lower case letter values in calligraphic letter alphabets. 
A stochastic kernel / transition probability from $\mathcal{X}$ to $\mathcal{Y}$ is denoted by $Q(y|x), \forall (x,y)\in \mathcal{X}\times\mathcal{Y}$, and the output marginal distribution induced by $P\in \mathcal{P}_{\mathcal{X}}$ through the channel $Q$ as $[PQ](y) := \sum_x P(x)Q(y|x), \forall  y\in\mathcal{Y}$ where $ \mathcal{P}_{\mathcal{X}}$ is the space of all distributions on $\mathcal{X}$.  As a shorthand notation, we also define $Q_{x^n}(.):=Q(.|x^n)$.
 We use $y_j^n :=[y_{j,1},..., y_{j,n}]$, and simply $y^n$ instead of $y_1^n$. 
The  empirical distribution of a sequence $x^n$ is %
\begin{align}
\widehat{P}_{x^n}(a)
:=\frac{1}{n}\mathcal{N}(a\vert x^n)=\frac{1}{n}\sum_{i=1}^n \mathbbm{1}_{\{x_i=a\}}, 
\forall a \in \mathcal{X},
\label{eq:empirical dist}
\end{align}
where $\mathbbm{1}_{\{A\}}$ is the indicator function of the event $A$ and where $\mathcal{N}(a\vert x^n)$ denotes the number of occurrences of letter $a\in \mathcal{X}$ in the sequence $x^n$;
when using~\eqref{eq:empirical dist} the target sequence $x^n$ is usually clear from the context so we may drop  the subscript $x^n$. The $P$-type set and the $V$-shell of the sequence $x^n$ are respectively  defined as 
\begin{align*}
T(P)&:=\left\{x^n: \mathcal{N}(a\vert x^n)=nP(a),\forall a\in \mathcal{X} \right\}\\
T_V(x^n)&:=\left\{y^n\!:\! \frac{\mathcal{N}\left(a,b\vert x^n, y^n \right)}{\mathcal{N}(a\vert x^n)}=V(b|a),\forall (a, b)\in (\mathcal{X},\mathcal{Y})\right\}\
\end{align*}
where $\mathcal{N}(a,b\vert x^n, y^n) =\sum_{i=1}^n \mathbbm{1}_{\left\{\substack{x_i=a\\y_i=b} \right\}}$ is the number of joint occurrences of $(a,b)$ in the pair of sequences $(x^n, y^n)$.
We also use $I(P,Q)$ to denote the mutual information between random variable {$(X,Y)\sim (P,[PQ])$ coupled via $P_{Y|X}(y|x)=Q(y|x)$}, 
$D(P_1 \parallel P_2)$ for the Kullback Leibler divergence between distribution $P_1$ and $P_2$, and $D(Q_1 \parallel Q_2|P) :=\sum_{x,y\in\mathcal{X}\times \mathcal{Y}} P(x)Q_1(y|x) \log \frac{Q_1(y|x)}{Q_2(y|x)}$ for conditional Kullback Leibler divergence.

\section{ System model for fixed number of transmissions and  main results}
\label{sec: sync+data}
We consider a discrete memoryless channel with transition probability matrix $Q(y|x)$ defined over all $(x,y)$ in the finite input and output alphabets $(\mathcal{X},\mathcal{Y})$. We also define a noise symbol $\star\in \mathcal{X}$ for which $Q_\star(y)>0,\ \forall y\in \mathcal{Y}$. 

An $(M, A, K, n, \epsilon)$ code for the {\it slotted bursty and strongly asynchronous} discrete memoryless channel 
with transition probability matrix $Q(y|x)$ {\it with fixed number of transmissions} consists of: 
\begin{itemize}
\item A message set $ [ M]$, from which messages are selected uniformly at random.
\item  Encoding functions  $f_i:  [M]\to \mathcal{X}^n, \ i\in  [A]$,  where we define $x_i^n(m):=f_i(m)$.  
The transmitter chooses uniformly at random one set of $K$ blocks for transmission out of the ${A \choose K}$ possible ones, and a set of $K$ messages from $M^{K}$ possible ones, also uniformly at random, and sends $x_{\nu_i}^n(m_i)$ in block $\nu_i$ for $i\in  [K]$ and $\star^n$ in every other block. We denote the chosen blocks and messages as $\left((\nu_1,m_1), \ldots, (\nu_{K},m_{K})\right)$.
\item A destination decoder function 
\[
g(\mathcal{Y}^{nA})= \left((\widehat{\nu}_1,\widehat{m_1}),\ldots, (\widehat{\nu}_K,\widehat{m_K})\right), 
\]
such that the average probability of error associated to it, given by
\begin{align*}
P_e^{(n)}:=\frac{1}{M^{K} \binom{A}{K}}\sum_{(\nu_1,m_1),\ldots,(\nu_{K},m_{K})}
\PP[g(y^{nA})\neq \left((\nu_1,m_1),\ldots,(\nu_{K},m_{K})\right)|H_{\left((\nu_1,m_1),\ldots,(\nu_{K},m_{K})\right)}],
\end{align*} 
satisfies $P_e^{(n)}\leq \epsilon,$ where $H_{\left((\nu_1,m_1),\ldots,(\nu_{K},m_{K})\right)}$ is the hypothesis that user transmits message $m_i$  at block $\nu_i$ with the codebook $x^n_{{\nu_i}}(m_i)$, for all $i\in [K]$.
\end{itemize}

A tuple $(R, \alpha, \nu)$ is said to be achievable if there exists a sequence of codes $(e^{nR}, e^{n\alpha}, e^{n\nu}, n,\epsilon_n)$ with $\epsilon_n$ going to zero  as $n$ goes to infinity.  The capacity region is the set of all possible achievable $(R, \alpha, \nu)$ triplets.  

We now introduce our main result. In Theorem~\ref{thm:ach R>0} we show that an exponential number of transmissions for a single user is possible at the expense of a reduced rate and/or reduced asynchronous window length compared to the case of only one transmission $K_n=1$ (or more generally $\nu =0$).  
\begin{theorem} 
\label{thm:ach R>0}
Achievable  and impermissible regions for the capacity region  
of a slotted bursty and strongly asynchronous discrete memoryless channel with transition probability matrix $Q(y|x)$ are given by
\begin{align}
\mathcal{R}^{in}
:=\bigcup_{\lambda\in[0,1], P\in \mathcal{P}_{\mathcal{X}}}&
\begin{Bmatrix}
{\nu\leq \alpha} \\
\alpha+R<D(Q_{\lambda} \parallel Q_\star|P)\\
\nu<D(Q_{\lambda} \parallel Q|P)\\
R<I(P,Q)\end{Bmatrix},
\label{eq:lb alpha nu R ach}
\end{align}
and
\begin{align}
\mathcal{R}^{out}
:=\bigcup_{\lambda\in[0,1], P\in \mathcal{P}_{\mathcal{X}}}&
\begin{Bmatrix}
\left\{\nu > \alpha\right\} \cup
\begin{Bmatrix}
\alpha>D([PQ_{\lambda}] \parallel Q_\star)+\left[ I(P,Q_\lambda)-R\right]^+\\
\nu>D(Q_{\lambda} \parallel Q|P)
\end{Bmatrix} \cup
\left\{R>I(P,Q)\right\}\end{Bmatrix},
\label{eq:lb alpha nu R conv}
\end{align}
where 
\begin{align}
Q_{\lambda}(.\vert x)&:=\frac{ Q_x^{\lambda}(.)Q_\star^{1-\lambda}(.)}{\sum_{y'\in \mathcal{Y}}Q_x^{\lambda}(y')Q_\star^{1-\lambda}(y')}\label{Eq:Q_lambda def}.
\end{align}
\end{theorem}
\begin{IEEEproof}
{\bf Achievability.} {\it Codebook generation.}
The user generates $A_n$ constant composition codebooks, of rate $R$ and blocklength $n$, by selecting each message's codeword uniformly and independently from the $P$-type set of sequences in $\mathcal{X}^n$, one codebook for each available block. 

{\it Decoder.} We perform a two-stage decoding. First, the decoder finds the location of the transmitted codewords (first stage, the synchronization stage) and it decodes the messages (second stage, the decoding stage). The probability of error for this two-stage decoder is given by
\begin{align*}
P_e^{(n)} 
& \leq 
\PP[\text{synchronization error}]+
\PP[\text{decoding error} | { \text{no synchronization error}}].
\end{align*}
For the first stage, fix  
\[T: -D(Q_\star\parallel Q|P)\leq T\leq D(Q\parallel Q_\star|P),\]
 which can be changed for different trade-off points. At each block $ j\in[A_n]$, if there exists {\it any} message $m\in [M_n]$ such that the Log Likelihood Ratio (LLR)
\begin{align}
L\left(y_j^n,x_j^n(m)\right):=\frac{1}{n}\log\frac{Q\left(y_j^n|x_j^n(m)\right)}{Q_{\star^n}(y_j^n)}\geq T,\label{eq:L}
\end{align}
 declare a codeword transmission block and a noise block otherwise. 
Given the hypothesis \[H_1:=H_{\left((1,1),\ldots,(K_n,1)\right)}\] the probability of the synchronization error in the first stage is given by
\begin{align}
&\PP\left[\text{synch error}\vert H_1\right]\notag\\
&\leq \prob\left[\bigcup_{j=1}^{K_n} \bigcap_{m=1}^{M_n}L\left(Y_j^n, x_j^n(m)\right)<T \vert H_1 \right]\notag\\
&\quad + \prob\left[\bigcup_{j=K_n+1}^{A_n}  \bigcup_{m=1}^{M_n}L\left(Y_j^n, x_j^n(m) \right)\geq T \vert H_1 \right]\notag\\
&\leq \sum_{j=1}^{K_n}\PP\left[L\left(Y_j^n, x_j^n(1)\right)<T \vert H_1\right]+e^{nR}\sum_{j=K_n+1}^{A_n} \PP\left[L\left(Y_j^n, x_j^n(1) \right)\geq T\vert H_1\right]\notag\\
&\leq e^{n\nu}\sum_{\substack{\hq:\\ D(\hq||Q_\star|P)-D(\hq||Q|P)<T}}\PP\left[Y^n\in T_{\hq}\left(x^n(1)\right) \vert H_1\right]\notag\\
&\quad +e^{n\left(R+\alpha\right)}\sum_{\substack{\hq:\\ D(\hq||Q_\star|P)-D(\hq||Q|P)\geq T}} \PP\left[Y^n\in T_{\hq}(x^n(1)) \vert H_1\right]\label{eq:min exp 1}\\
&\leq  e^{n\nu}e^{-nD(Q_{\lambda} \parallel Q|P)}+e^{n(\alpha+R)} e^{-nD(Q_{\lambda} \parallel Q_\star|P)},\label{eq:optim prob}
\end{align}
where $Q_\lambda$ is defined in~\eqref{Eq:Q_lambda def} and 
\[
\lambda  : D(Q_{\lambda } \parallel Q_\star|P)-D(Q_{\lambda } \parallel Q|P)=  T 
.\]
 The expression in~\eqref{eq:optim prob} is the result of finding the minimum exponent in~\eqref{eq:min exp 1} using the Lagrangian method as in~\cite[Sec. 11.7]{coverbook}.

By~\eqref{eq:optim prob}, the probability of error in the synchronization goes to zero as $n$ goes to infinity when
\begin{subequations}
\begin{align}
\nu&<D(Q_{\lambda} \parallel Q\vert P),\\
\alpha+R&<D(Q_{\lambda} \parallel Q_\star\vert P).\label{eq:achievable alpha R}
\end{align}
\end{subequations}

Conditioning on the `no synchronization error' and  having found all $K_n$ `not noisy' blocks, we can use a Maximum Likelihood (ML) decoder for random constant composition codes, introduced and analyzed in~\cite{moulin2012log}, on the super-block of length $nK_n$ to distinguish among $e^{nK_nR}$ different message combinations. If $R < I(P, Q)$, the  probability of the error of the second stage also vanishes as $n\to \infty$. 

{\bf Converse.} The main technical difficulty and innovation in the proof relies on analyzing the probability of error in a ML decoder. 
In this regard, we boil down the problem to finding an exponentially decaying `lower' bounds on the probability of the missed detection (where the likelihood ratio defined in~\eqref{eq:L} of an active block  is less than a threshold) and false alarm (where the likelihood ratio defined in~\eqref{eq:L} of an idle block is larger than the threshold) error events. 
By the type counting argument and the fact that we have polynomially many types in blocklength at the expense of a small reduce in rate~\cite{csiszarbook} we can restrict our attention to constant composition codes. In other words, we assume the use of  codewords $x_i^n(.)$ with  constant compositions $P_{i}$ in each block $i\in [A_n]$.  
Given the hypothesis $H_1:=H_{((1,1)\ldots(K_n,1))}$, with a ML decoder (which achieves the minimum average probability of error) and for any $T\in \mathbb{R}$, the error events are given by
\begin{align}
  &\left\{\text{error}|H_1\right\}=\bigcup_{\substack{
\left((l_1,\tilde{m}_1 )\ldots, (l_{K_n},\tilde{m}_{K_n}) \right)
\\ \neq 
 \left((1,{m}_1),\ldots, (K_n,{m}_{K_n}) \right)}}
\left\{
     \sum_{i=1}^{K_n} L\left(Y_i^n,x_{  i}^n(m_i)\right) 
\leq \sum_{i=1}^{K_n} L\left(Y_{l_i}^n,x_{  l_i}^n(\tilde{m}_i)\right)
\right\}
\label{eq:all errors conv},
\end{align}
where~\eqref{eq:all errors conv} the union of the events that the sum of the LLRs of the true hypothesis $ \left((1,{m}_1),\ldots, (K_n,{m}_{K_n}) \right)$ is less than the sum of the LLRs of the wrong hypothesis (with arbitrary number of incorrect synchronization or decoding errors)
\[\left((l_1,\tilde{m}_1 )\ldots, (l_{K_n},\tilde{m}_{K_n}) \right)\neq 
 \left((1,{m}_1),\ldots, (K_n,{m}_{K_n}) \right),\]
 where wrong means that we have at least one decoding or one synchronization error. We now focus our attention on a subset of these events which have a {\it single} synchronization error. i.e., 
\begin{align}
&\left\{\text{error}|H_1\right\}\supseteq \bigcup_{\substack{i\in [K_n]\\j\in[K_n+1:A_n]\\{m} \in [M_n]}} 
\big\{
L\left(Y_i^n,x_{  i}^n(m_i) \right)\leq L\left (Y_j^n, x_{ j}^n({m})\right) 
\big\}
\label{eq:event lower bound}
\\&\supseteq
\left\{
\bigcup_{i\in[K_n]}
\left\{L\left(Y_i^n,x_{  i}^n(m_i)\right)\leq T  \right\}
\right\}
\bigcap 
\left\{
\bigcup_{\substack{ j\in[K_n+1:A_n]\\m\in[M_n]}} 
\left\{ L\left(Y_j^n,x_{ j}^n(m)\right)\geq T  \right\}
\right\}.
\label{eq:main prob}
\end{align}
In other words,~\eqref{eq:event lower bound} is  the union over the events that  {\it (any message, any noisy block)} is selected instead of {\it one} of the {\it (correct message, correct block)}s; with the underlying assumption that the rest of the blocks are chosen correctly. We also further restrict 
\[T\in [-D(Q_\star\parallel Q|P_{i^\star}),D(Q\parallel Q_\star |P_{i^\star}) ],\]
 where $i^\star$ is chosen such that 
\begin{align}
i^\star:=\arg\max_{\substack{ i,\lambda_i:\\ D(Q_{\lambda_i} \parallel Q_\star|P_{i })- D(Q_{\lambda_i}\parallel Q|P_{i })=T }}\hspace{-.6cm} D(Q_{\lambda_i}\parallel Q|P_i)
.\label{eq:i^star}
 \end{align}
 
 The reason for this choice of $i^*$ will be become clear later (see~\eqref{eq:first lower bound} and~\eqref{eq:second lower bound}).
 By~\eqref{eq:main prob} we have
\begin{align}
&\prob \left[\text{error} \Big|H_1\right]\notag
\\
&\geq 
\prob \left[
\bigcup_{i\in [K_n]} L\left (Y_i^n, x_{  i}^n({m})\right) \leq T \vert H_1
 \right]  
\cdot 
\prob \left[ 
\bigcup_{\substack{j\in[K_n+1:A_n]\\m\in [M_n]}}  L\left (Y_j^n, x_{  j}^n({m})\right) \geq T \vert H_1
 \right]
\label{eq:err lb 2}\\
&\geq \left( 1-e^{-n\left[ \nu - D\left(Q_{\lambda_{i^\star}} \parallel Q|P_{i^\star} \right) \right]} \right)
 \label{eq:first lower bound}\\
&\quad \cdot\left( 1-e^{-n\left[ \alpha+R\mathbbm{1}_{\left\{R<I(P, Q_{\lambda_{i^*}}) \right\}}-D(Q_{\lambda_{i^*}}\parallel Q_\star|P_{i^*})\right]}\right),
\label{eq:second lower bound}
\end{align}
where~\eqref{eq:err lb 2} is due to the independence of $Y_j^n , j\in  [A_n]$ and where~\eqref{eq:first lower bound} and~\eqref{eq:second lower bound} are proved in Appendix~\ref{app:prob union less than T} and~\ref{app:union prob messages}, respectively. 

The lower bound on the probability of error given in~\eqref{eq:first lower bound} and~\eqref{eq:second lower bound}, would be bounded away from zero if
\begin{subequations}
\begin{align*}
\nu &> D\left(Q_{\lambda_{i^\star}} \parallel Q|P_{i^\star} \right),\\
\alpha+R\mathbbm{1}_{\left\{R<I(P,Q_{\lambda_{i^*}}) \right\}} &> D\left( Q_{\lambda_{i^\star}} \parallel Q_\star|P_{i^\star}\right)= I(P,Q_{\lambda_{i^\star}} )+D([ P_{i^*}Q_{\lambda_{i^\star}} ]\parallel Q_\star),
\end{align*}
\end{subequations}
which can be equivalently be written as
\begin{subequations}
\begin{align}
\nu &> D\left(Q_{\lambda_{i^\star}} \parallel Q|P_{i^\star} \right),\\
\alpha &> D([ P_{i^*}Q_{\lambda_{i^\star}} ]\parallel Q_\star)+\left[ I(P,Q_{\lambda_{i^\star}} )-R\right]^+,\label{eq:converse alpha R}
\end{align}\label{eq:converse R=0 impermissible}
\end{subequations}
and hence this region is impermissible.
 
Any asynchronous channel can be reduced to a synchronous channel by providing the decoder with  side information about the transmission time. Hence, the same bound on the rate of a synchronous  channel, i.e. $R<I(P_{i^\star},Q)$ also applies to the asynchronous channel. 
By the symmetry of the hypothesis, the same lower bound on probability of error holds for the average probability of error and hence we retrieve the bounds given in~\eqref{eq:lb alpha nu R conv}.
\end{IEEEproof} 
{ \begin{remark}
We note that for the set of $\lambda: R<I(P,Q)$, the achievability and converse bounds match. As the result, there is a region for $R$ small enough, that our achievability and converse bounds coincide. 
\end{remark}}
{The main novelty in this problem is to find exponentially decaying upper and lower bounds on the probability of error. The achievability scheme analysis is easier as we can easily pose it as a hypothesis testing problem. However, in the converse, we have to deal with the optimal ML decoder. As a first step in reducing the complexity of the ML decoder, we considered a set of error events with single synchronization errors (which we believe is the major error set and many other events are its subsets). Next, we had to find  the probability that the LLR's of the active blocks are smaller than a threshold. This again, would be easy to calculate for a single LLR; its probability is a function of the (imaginary) channel $Q_\lambda$ defined in~\eqref{Eq:Q_lambda def}. However, we have to deal with unions of such events as in~\eqref{eq:err lb 2}. Calculation of these unions is also easy for $\nu = 0$. In this case the optimal $\lambda =1 $ and hence $Q_{\lambda =1}=Q$ and one can leverage the fact that the probability of decoding error  for channel $Q$ is small to transform the union into a summation. If however $\nu\neq 0$ and hence $\lambda \neq 1$, probability of error for channel $Q_\lambda$ (for the same code as channel $Q$) would be dependent on the rate $R$. Transformation of a union to a summation is not straightforward anymore and hence we had to provide several additional steps (in Appendix~\ref{app:union prob messages} and~\ref{app:lower bound union messages}) to do so.}
 
\begin{figure}
\centering
\includegraphics[width = .5\textwidth]{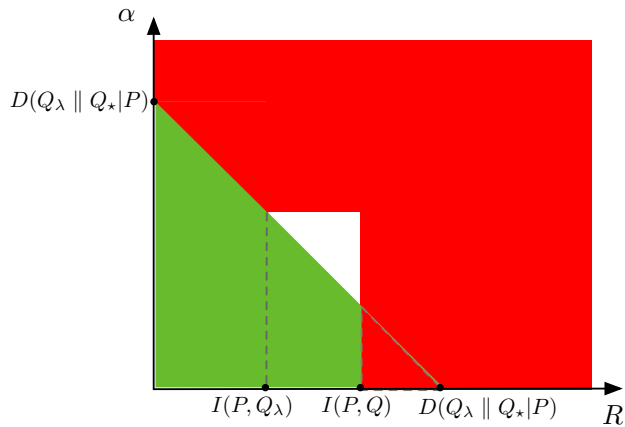}
\caption{Comparison of Impermissible region given in~\eqref{eq:converse alpha R} (red region) and achievable region given by~\eqref{eq:achievable alpha R} (green region) for fixed $\lambda$.}
\label{fig:conv ach alpha R}
\end{figure}

\begin{figure}[htbp]
 \centering
 \subfigure[Impermissible region]{\includegraphics[width=0.45\textwidth]{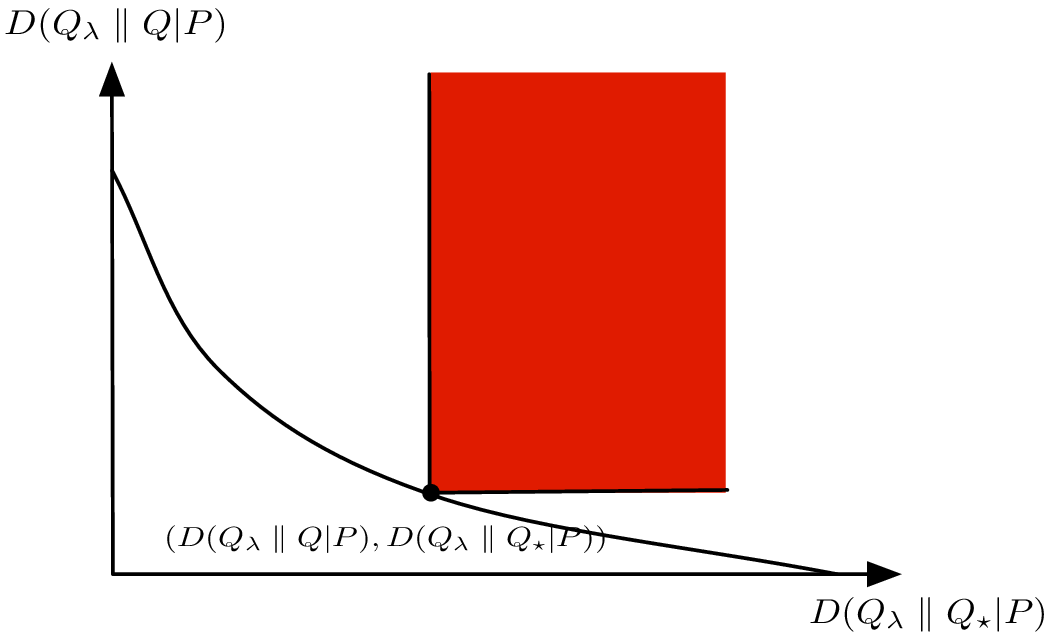}\label{fig:impermissible}}
 \subfigure[Achievable region]{\includegraphics[width=0.45\textwidth]{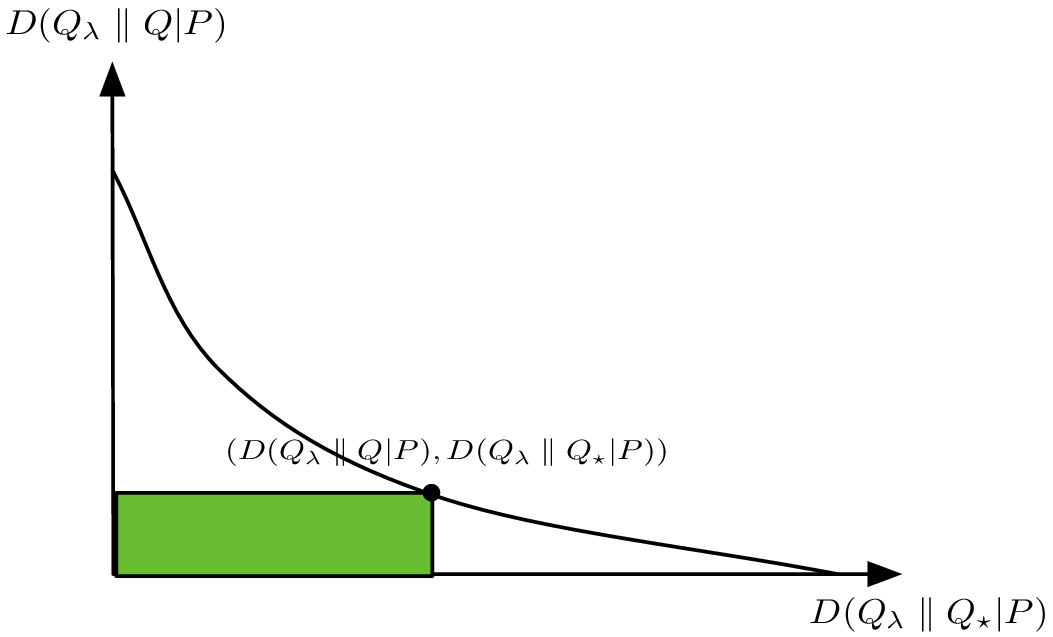}\label{fig:achievable}}
\caption{The union of the regions over different values of $\lambda$ will result in matching achievability and converse bounds for $R=0$.}
\label{fig:ach conv match}
 \end{figure}

For a fixed $\lambda$, a comparison between the bounds given by~\eqref{eq:converse alpha R} and~\eqref{eq:achievable alpha R} is shown in Fig.~\ref{fig:conv ach alpha R}. 
 It is easy to see that the bounds given in~\eqref{eq:lb alpha nu R ach} and~\eqref{eq:lb alpha nu R conv} will coincide (i.e., complement one another) for the case $\nu =0$ ($\lambda =1$) and retrieve the capacity region previously derived in~\cite{Strong_P2P-it}. 
 \begin{remark}
It is worth noting that the region specified in~\ref{fig:conv ach alpha R}, need not be convex since $\alpha$ is a channel parameter and can not be chosen by user.
\end{remark}
 We now concentrate our attention to the synchronization case only. 
\begin{remark}\label{rmk:R=0}
 By specializing Theorem~\ref{thm:ach R>0} for $R=0$, we can see that $\mathcal{R}^{in}\vert_{R=0}=\mathcal{R}^{out}\vert_{R=0}=\mathcal{R}\vert_{R=0}$.
  \end{remark}
  It can be easily seen in Fig.~\ref{fig:ach conv match} that by taking the union over $\lambda\in [0,1]$, the achievability and converse regions match for $R=0$.

In the following example, we consider a Binary Symmetric Channel and plot its achievable region.
\begin{example}\label{ex:BSC}
To illustrate the capacity region in Theorem~\ref{thm:ach R>0}, we consider a Binary Symmetric Channel (BSC) $Q$ with cross over probability $\delta $ as it is shown in Fig.~\ref{fig:BSC}. We also assume $\star =0$. For the channel $Q_\lambda$ in~\eqref{Eq:Q_lambda def} we have
\begin{align*}
Q_\lambda(0\vert 0)&=1-\delta,\\
\epsilon_\lambda:=Q_\lambda(0\vert 1)&=\frac{\delta^\lambda (1-\delta)^{(1-\lambda)}}{\delta^\lambda (1-\delta)^{(1-\lambda)}+(1-\delta)^\lambda \delta^{(1-\delta)}}.
\end{align*}

\begin{figure}[htbp]
\centering
\includegraphics[width=.8\textwidth]{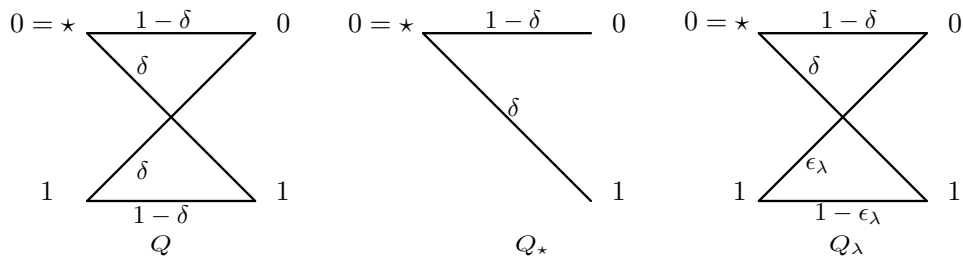}
\caption{Strongly synchronous binary symmetric channel}
\label{fig:BSC}
\end{figure}
By changing $p=\PP[X=0]\in [0,\frac{1}{2}]$ and $\lambda \in [0,1]$, we obtain the achievability region shown in Fig.~\ref{fig:BSC cap region}. In addition, the (optimal) trade off for $(R, \alpha, \nu=0)$ can be seen in Fig.~\ref{fig:BSC R alpha} which resembles the one in~\cite[Fig. 1]{polyanskiy}. The trade off between $(\alpha, \nu)$ can be seen in Fig.~\ref{fig:BSC alpha nu} which has the curvature we expect to see, like the one in Fig.~\ref{fig:l_a l_star} in the Appendix. 
 \begin{figure}[htbp]
 \begin{minipage}[t]{0.55\textwidth}
\vspace{2cm}
 \subfigure[$(R,\alpha, \nu)$ trade-off]{\includegraphics[width=1\textwidth]{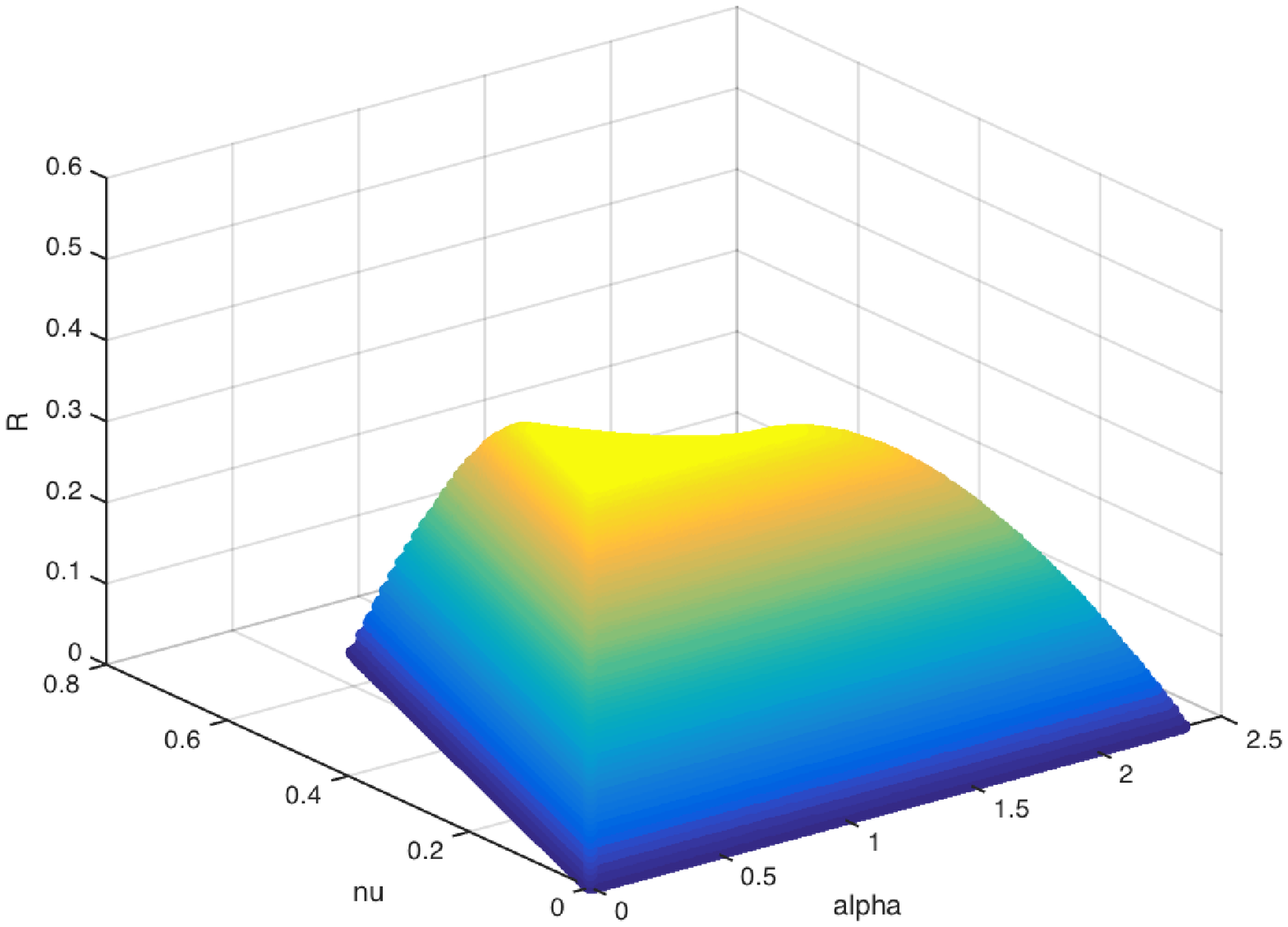} \label{fig:BSC cap region}}
 \end{minipage}
 \hspace*{\fill}
 \begin{minipage}[t]{0.45\textwidth}
 \subfigure[$(R, \alpha)$ trade-off for $\nu=0$]{\includegraphics[width=1\textwidth]{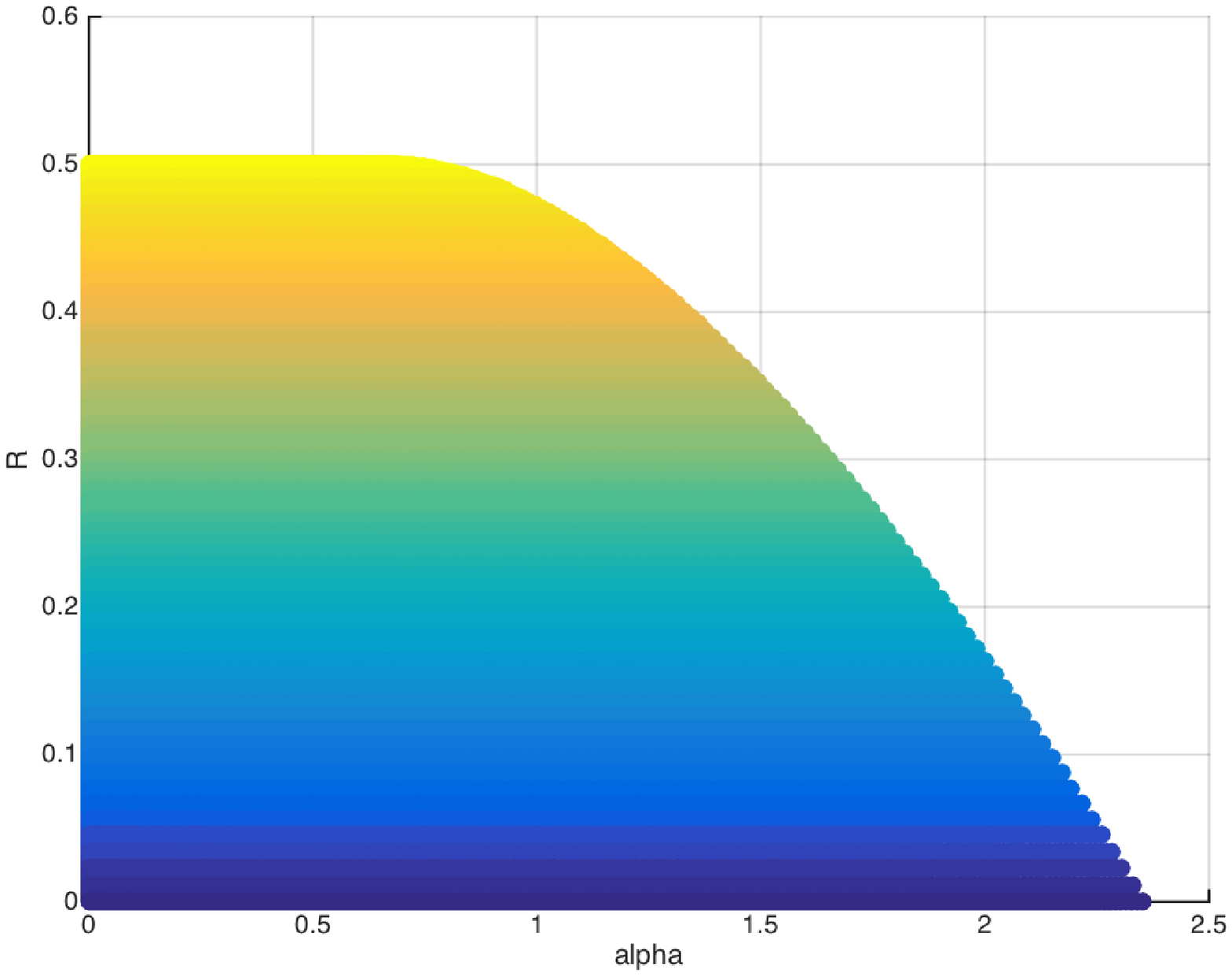}\label{fig:BSC R alpha}}\\
\subfigure[$(\alpha, \nu)$ trade-off for different rates, specified by the color]{\includegraphics[width=1\textwidth]{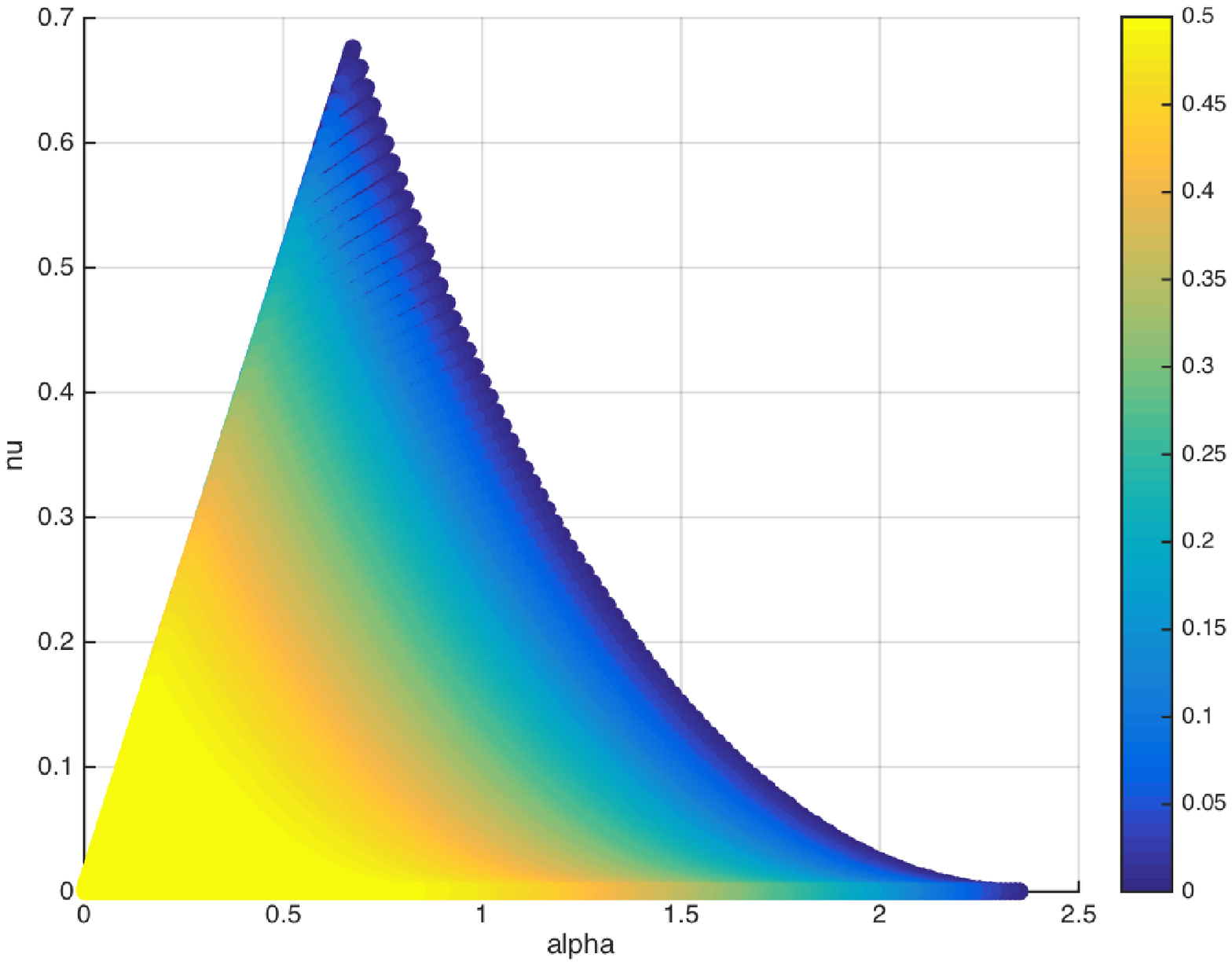}\label{fig:BSC alpha nu}}
 \label{fig:BSC 3 figs}
  \end{minipage}
\caption{Achievability bound on capacity region of slotted bursty and strongly asynchronous BSC with fixed number of transmissions with cross over probability $\delta =0.11$.}
 \end{figure}
  \end{example}
 
Theorem~\ref{eq:equiv thm for R=0} provides another form for the trade-off between $(R=0, \alpha, \nu)$ which implies that using a repetition pattern for synchronization pattern is optimal.
\begin{theorem}\label{eq:equiv thm for R=0}
For $R=0$, the capacity region $\mathcal{R}\vert_{R=0}$ in Remark~\ref{rmk:R=0}  is equivalent to
\begin{align}
\mathcal{R}^{\text{synch}}:=\bigcup_{x\in\mathcal{X},\lambda\in[0,1]}&
\begin{Bmatrix}
 \nu< \alpha \\
 \alpha <D(Q_{\lambda} \parallel Q_\star) \\ 
 \nu    < D(Q_{\lambda}  \parallel Q_x)
\end{Bmatrix}.
\label{eq:main R=0}
\end{align}
\end{theorem}
\begin{IEEEproof}
 $\mathcal{R}^{\text{synch}} \subseteq \mathcal{R}|_{R=0}$ is trivial since we can restrict the set of distributions $P\in \mathcal{P}_{\mathcal{X}}$ in $\mathcal{R}|_{R=0}$ to the distributions with weight one on a single symbol $x$ and zero weight on all other symbols. 
 
We also prove $\mathcal{R}|_{R=0} \subseteq \mathcal{R}^{\text{synch}}$ by contradiction and by means of the following Lemma proved in Appendix~\ref{app:region}.
\begin{lemma}\label{lemma:lower envelope}
The curve $\left( D(Q_\lambda\parallel Q_\star|P),  D(Q_\lambda\parallel Q|P)\right)$  characterized by $\lambda\in[0,1]$ is the lower envelope of the set of curves
 \[\bigcup_{x\in \mathcal{X}}\big\{\left( D(Q_{\lambda_x}\parallel Q_\star|P),  D(Q_{\lambda_x}\parallel Q|P)\right)\big\},\] 
 which are each characterized by $\lambda_x\in [0,1]$.
\end{lemma}

We continue the proof by assuming $ \mathcal{R}|_{R=0}\not \subseteq \mathcal{R}^{\text{synch}}$. Then there exists an element 
\begin{align*}
(r_1,r_2,0) &= \left(D(Q_{\lambda}\parallel Q|P),
D(Q_{\lambda}\parallel Q|P ),0 \right)\in \mathcal{R}|_{R=0},\\
 (r_1,r_2)  &\not\in \mathcal{R}^{\text{synch}},
\end{align*}
 that is,  
which lies above all the $\left\{D(Q_{\lambda}\parallel Q_\star), D(Q_{\lambda}\parallel Q_x)\right\}$ curves for all $x\in \mathcal{X}$. 
 Hence, for any $x\in \mathcal{X}$, there exists a $\lambda_x$ such that
 \begin{align*}
r_1=D(Q_{\lambda}\parallel Q|P)>D(Q_{\lambda_x}\parallel Q_\star),\\
r_2=D(Q_{\lambda }\parallel Q|P ) > D(Q_{\lambda_x}\parallel Q_x).
 \end{align*}
As a result
  \begin{align*}
D(Q_{\lambda}\parallel Q|P)>D(Q_{\lambda_x}\parallel Q_\star|P),\\
D(Q_{\lambda}\parallel Q|P ) > D(Q_{\lambda_x}\parallel Q|P),
 \end{align*} 
 which contradicts  Lemma~\ref{lemma:lower envelope} that $\left(D(Q_{\lambda}\parallel Q|P),D(Q_{\lambda}\parallel Q|P )  \right)$ is the lower envelope of the set of $\bigcup_{x\in \mathcal{X}}\left\{ \left(D(Q_{\lambda_x}\parallel Q_\star|P),D(Q_{\lambda_x}\parallel Q|P)\right)\right\}$ curves and hence the initial assumption that $\mathcal{R}|_{R=0}\not \subseteq \mathcal{R}^{\text{synch}}$ is not feasible.
 \end{IEEEproof}
 {  Note that by adapting the achievability scheme to synchronize only ($R=0$), 
we do not need  a different  synchronization pattern for each  block. Using the same synchronization 
  pattern in every block suffices to drive the probability of error in the synchronization stage to zero and since it matches the converse, it is optimal.}

\begin{figure}[htbp]
  \centering
 \subfigure[]{\includegraphics[height=0.4\textwidth]{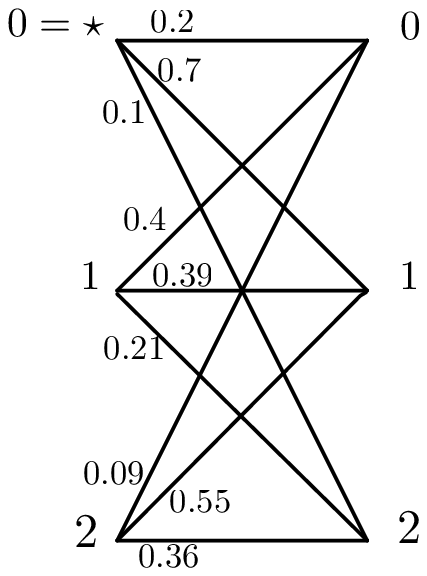}\label{fig:ternary_channel}}
 \subfigure[]{\includegraphics[height=0.5\textwidth]{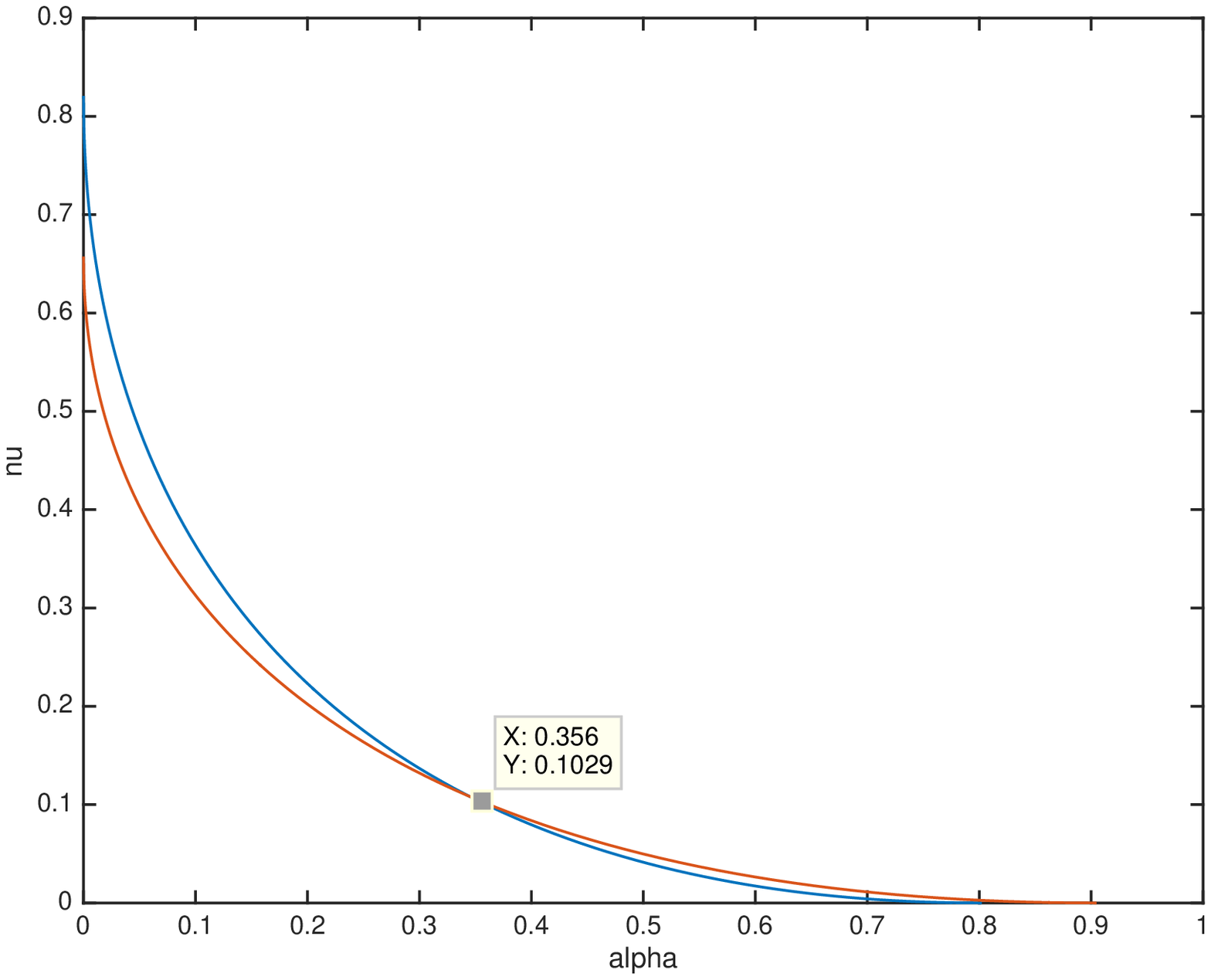}\label{fig:ternary_symbols}}
\caption{Channel with different synchronization pattern symbols for different $(\alpha, \nu)$ regimes.}
\end{figure}

  Theorem~\ref{eq:equiv thm for R=0} also implies that depending on the value of $\alpha$ and $\nu$, using a repetition synchronization pattern with a single symbol is optimal. This symbol may change depending on the considered value of $\alpha$ and $\nu$. For the ternary channel in Fig.~\ref{fig:ternary_channel}, for example, the resulting curves by using symbol $x=1$ and $x=2$ are shown in Fig.~\ref{fig:ternary_symbols}. As it is clear, for the regime $\alpha>0.356$, symbol $x=1$ has to be used whereas in the regime $\alpha\leq 0.356$ symbol $x=2$ has to be used in the synchronization pattern.

 \section{System model for random transmissions and  main result}\label{sec:random transmission}
We consider again a discrete memoryless channel with transition probability matrix $Q(y|x)$ defined over all $(x,y)$ in the finite input and output alphabets $(\mathcal{X},\mathcal{Y})$. We also define a noise symbol $\star\in \mathcal{X}$ for which $Q_\star(y)>0,\ \forall y\in \mathcal{Y}$. 

An $(M, A, p, n, \epsilon)$ code for the {\it slotted bursty and strongly asynchronous} discrete memoryless channel 
with transition probability matrix $Q(y|x)$ 
{\it with random access} is defined as follows. 
\begin{itemize}
\item A message set $[M]$, from which messages are selected uniformly at random.
\item Encoding functions  $f_i:  [M]\to \mathcal{X}^n, \ i\in  [A]$, 
where we define $x_i^n(m):=f_i(m)$.  
For each block $i\in  [A]$, the transmitter chooses a message among $M$ possible ones and transmit $x_i^n(m_i)$ through the channel with probability $p$ or remains idle and transmits $\star^n$ with probability $1-p$.  
\item A destination decoder function
\[
g(\mathcal{Y}^{nA})= \left((\widehat{\nu}_1,\widehat{m_1}),\ldots, (\widehat{\nu_{\hat{k}}},\widehat{m_{\hat{k}}})\right),
\]
such that the average probability of error associated to it, given by
\begin{align*}
P_e^{(n)}:=&\sum_{k=1}^A \sum_{(\nu_1,m_1),\ldots,(\nu_k,m_{k})}\frac{1}{M^{k} }p^k(1-p)^{A-k}
\PP[g(y^{nA})\neq \left((\nu_1,m_1),\ldots,(m_{k},\nu_{k})\right)|H_{\left((\nu_1,m_1),\ldots,(\nu_k,m_{k})\right)}],
\end{align*}
satisfies $P_e^{(n)}\leq \epsilon,$ where $H_{\left((\nu_1,m_1  ),\ldots,(\nu_{k},m_{k} )\right)}$ is the hypothesis that user transmits message $m_i$  at block $\nu_i$ with the codebook $x^n_{{\nu_i}}$, for all $i\in  [k]$.
\end{itemize}
A tuple $(R, \alpha,\beta)$ is said to be achievable if there exists a sequence of codes $(e^{nR}, e^{n\alpha},  e^{-n\beta}, n,\epsilon_n)$ with $\epsilon_n\to 0$ as $n\to \infty$.  The capacity region is the set of all possible achievable $(R, \alpha, \beta)$ triplets.  

\begin{theorem}\label{thm: random access}
Achievable  and impermissible regions for  the capacity region  
of a slotted bursty and strongly asynchronous  random access  channel with transition probability matrix $Q(y|x)$ are given by
\begin{align}
\mathcal{R}^{in}
:=\bigcup_{\lambda\in[0,1], P\in \mathcal{P}_{\mathcal{X}}}&
\begin{Bmatrix}
\alpha+R&<D(Q_{\lambda} \parallel Q_\star|P)\\
\alpha-\beta&<D(Q_{\lambda} \parallel Q|P)\\
R&<I(P,Q)
\end{Bmatrix},
\label{eq:bound a and a+R ach}
\end{align}
and
\begin{align}
\mathcal{R}^{out}
:=\bigcup_{\lambda\in[0,1], P\in \mathcal{P}_{\mathcal{X}}}&
\begin{Bmatrix}
\begin{Bmatrix}
\alpha&>D([PQ_{\lambda}] \parallel Q_\star)+\left[I(P,Q_\lambda)-R \right]^+\\
\alpha-\beta&>D(Q_{\lambda} \parallel Q|P)
\end{Bmatrix}\cup
 \left\{ R>I(P,Q)\right\}
\end{Bmatrix}.
\label{eq:bound a and a+R conv}
\end{align}
\end{theorem}

\begin{IEEEproof}
{\bf Achievability. }
The encoder and decoder are the same as the one given for the achievability proof of Theorem~\ref{thm:ach R>0}, except that the number of active blocks is not fixed. We denote $p_n:=e^{-n\beta}$ and $\hat{H}_k$ to be the hypothesis that the user is active in $k$ blocks. 
By the symmetry of the probability of error among hypotheses with the same number of occupied blocks, we can write
\begin{align}
P_e^{(n)} &= \sum_{k=0}^{A_n}\binom{A_n}{k}p_n^k (1-p_n)^{A_n-k} \ \PP[\text{Error}|\hat{H}_k]
\notag\\
&\leq
 \sum_{k=0}^{A_n}\binom{A_n}{k}p_n^k (1-p_n)^{A_n-k}  \PP[\text{Synchronization error}|\hat{H}_k]\label{eq:sync error v2}
\\&\quad +
\sum_{k=0}^{A_n}\binom{A_n}{k}p_n^k (1-p_n)^{A_n-k}  \PP[\text{Decoding error} | \hat{H}_k,{ \text{No synchronization error}}].\notag
\end{align}
 With similar steps as those in the proof of Theorem~\ref{thm:ach R>0}, we obtain
 \begin{align}
 \PP[\text{synchronization error}|\hat{H}_k] & 
\leq
  k \ e^{-nD(Q_{\lambda} \parallel Q|P)}+e^{nR}\left(e^{n\alpha}-k\right)  e^{-nD(Q_{\lambda} \parallel Q_\star|P)},\label{eq:optim prob v2}
 \end{align}
where
\[
\lambda  : D(Q_{\lambda } \parallel Q_\star|P)-D(Q_{\lambda } \parallel Q|P)=  T .\]
 By~\eqref{eq:optim prob v2}, we can upper bound~\eqref{eq:sync error v2} as
\begin{align*}
& \sum_{k=0}^{A_n}\binom{A_n}{k}p_n^k (1-p_n)^{(A_n-k)}  \PP[\text{synchronization error}|\hat{H}_k]
 \\& \leq e^{n\alpha} e^{-n\beta} e^{-nD(Q_{\lambda} \parallel Q|P)} + e^{n(\alpha+R)} e^{-nD(Q_{\lambda} \parallel Q_\star|P)},
 \end{align*}
 which goes to zero for 
 \begin{align*}
 \alpha- \beta&<D(Q_{\lambda} \parallel Q|P),\\
 \alpha+R&<D(Q_{\lambda} \parallel Q_\star|P).
 \end{align*}
For the decoding stage, with the same strategy as the one in Theorem~\ref{thm:ach R>0} we obtain the third bound in~\eqref{eq:bound a and a+R ach}.

{\bf Converse. }
The converse argument is also similar to the converse proof of Theorem~\ref{thm:ach R>0}. It can be shown that
\begin{align}
\prob \left[\text{error} \Big|\hat{H}_k\right]
\geq&\left( 1-\frac{e^{ D\left(Q_{\lambda_{i^\star}} \parallel Q|P_{i^\star} \right)}}{k} \right)
\cdot
\left( 1-\frac{e^{-n\left[ R\mathbbm{1}_{\left\{R<I(P,Q_{\lambda_{i^*}}) \right\}} - D\left( Q_{\lambda_{i^\star}} \parallel Q_\star|P_{i^\star}\right) \right]}}{A_n-k}\right).\notag
\end{align}
 Hence
 \begin{align}
 \PP[\text{error}]&\geq  \sum_{k=1}^{A_n-1}\binom{A_n}{k}(e^{-n\beta})^k (1-e^{-n\beta})^{A_n-k} \left(1- \frac{e^{nD_1}}{k} \right)\left(1- \frac{e^{n\left(D_2-R\mathbbm{1}_{\left\{R<I(P,Q_{\lambda_{i^*}}) \right\}}\right)}}{A_n-k}  \right) 
 \notag
 \\&\geq 1-(1-e^{-n\beta})^{A_n}-e^{-n\beta A_n} -\frac{2e^{nD_1}}{e^{-n\beta} e^{n\alpha}}-\frac{2e^{n\left(D_2-R\mathbbm{1}_{\left\{R<I(P,Q_{\lambda_{i^*}}) \right\}}\right)}}{(1-e^{-n\beta})e^{n\alpha}}\label{eq:conv p lb},
 \end{align}
 where
 \begin{align*}
 D_1 & :=  D\left(Q_{\lambda_{i^\star}} \parallel Q|P_{i^\star} \right),\\
 D_2 & :=  D\left(Q_{\lambda_{i^\star}} \parallel Q_\star|P_{i^\star} \right),
 \end{align*}
and where~\eqref{eq:conv p lb} is proved in Appendix~\ref{app:conv p lb calculation}.
This retrieves the first two bounds in~\eqref{eq:bound a and a+R conv}. The third bound in~\eqref{eq:bound a and a+R conv} is by the usual bound on the reliable rate of a synchronous channel.
\end{IEEEproof}
It is easy to see that~\eqref{eq:bound a and a+R ach} and~\eqref{eq:bound a and a+R conv} match for the cases that $R=0$ or $\beta=\alpha$. The latter case corresponds to $\lambda = 1$.
\begin{example}
We consider the same BSC channel defined in Example~\ref{ex:BSC} and illustrate its achievability region for the slotted bursty and strongly asynchronous channel with random access in Fig.~\ref{fig:BSC random access CAP}. For  values of $\beta>D(Q\parallel Q_\star|P)=2.3527$, the achievable region is similar to the to the case $\beta=2.3527$ and the surface remains unchanged. This is also apparent in Fig.~\ref{fig:BSC random access alpha beta} where the trade-off between $(\alpha, \beta)$ is depicted. This is in fact obvious in Theorem~\ref{thm: random access} since for values of $\beta>D(Q\parallel Q_\star|P)$ the achievability~\eqref{eq:bound a and a+R ach} and converse bound~\eqref{eq:bound a and a+R conv} match and are equal to the capacity region for one for only one transmission as the one in~\cite[Fig. 1]{polyanskiy}.
 \begin{figure*}
 \subfigure[$(R,\alpha, \beta)$ trade-off]{\includegraphics[width=0.5\textwidth]{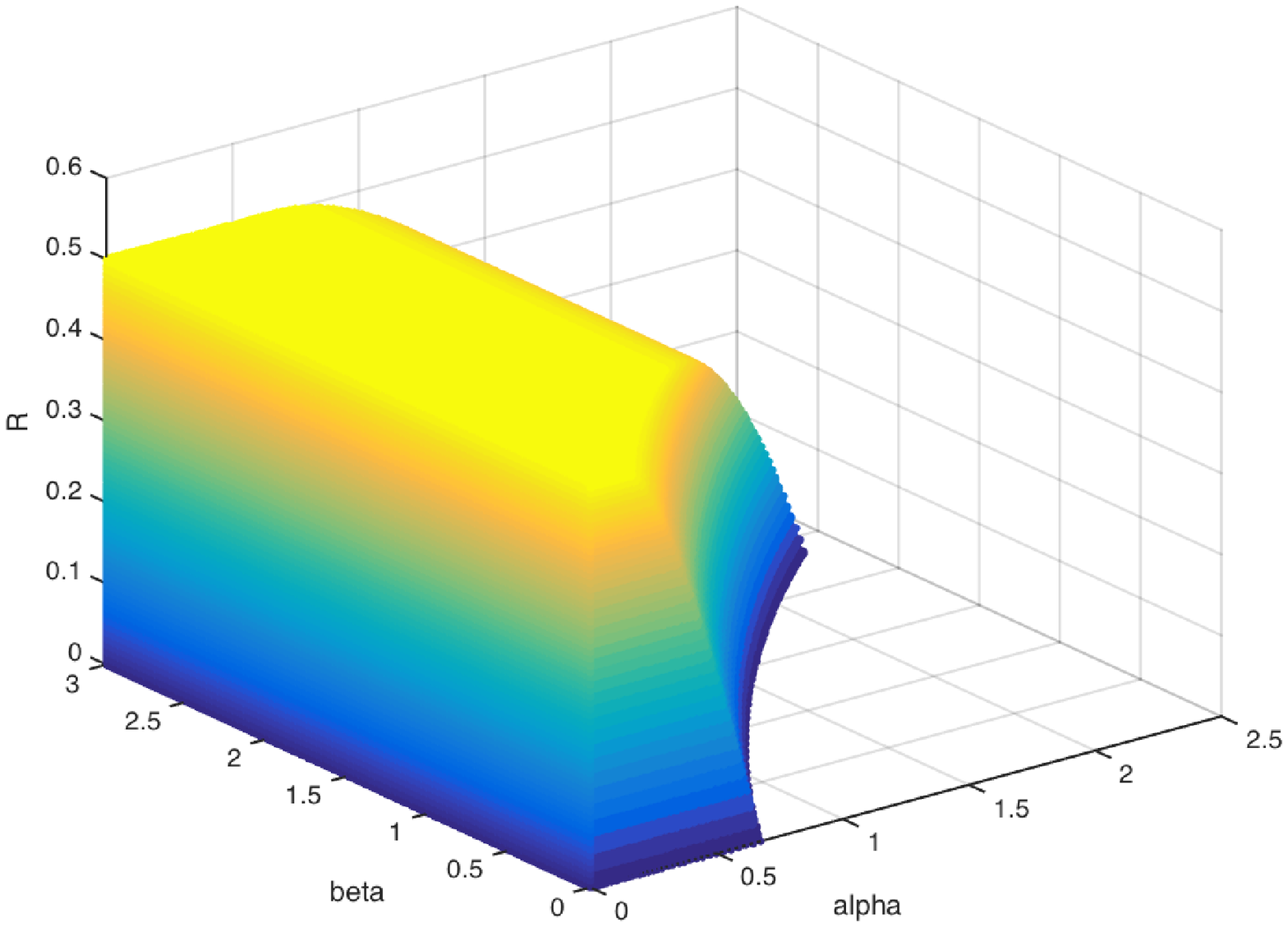}\label{fig:BSC  random access CAP}}
 \subfigure[$(\alpha, \beta)$ trade-off for different rates, specified by colors]{\includegraphics[width=0.5\textwidth]{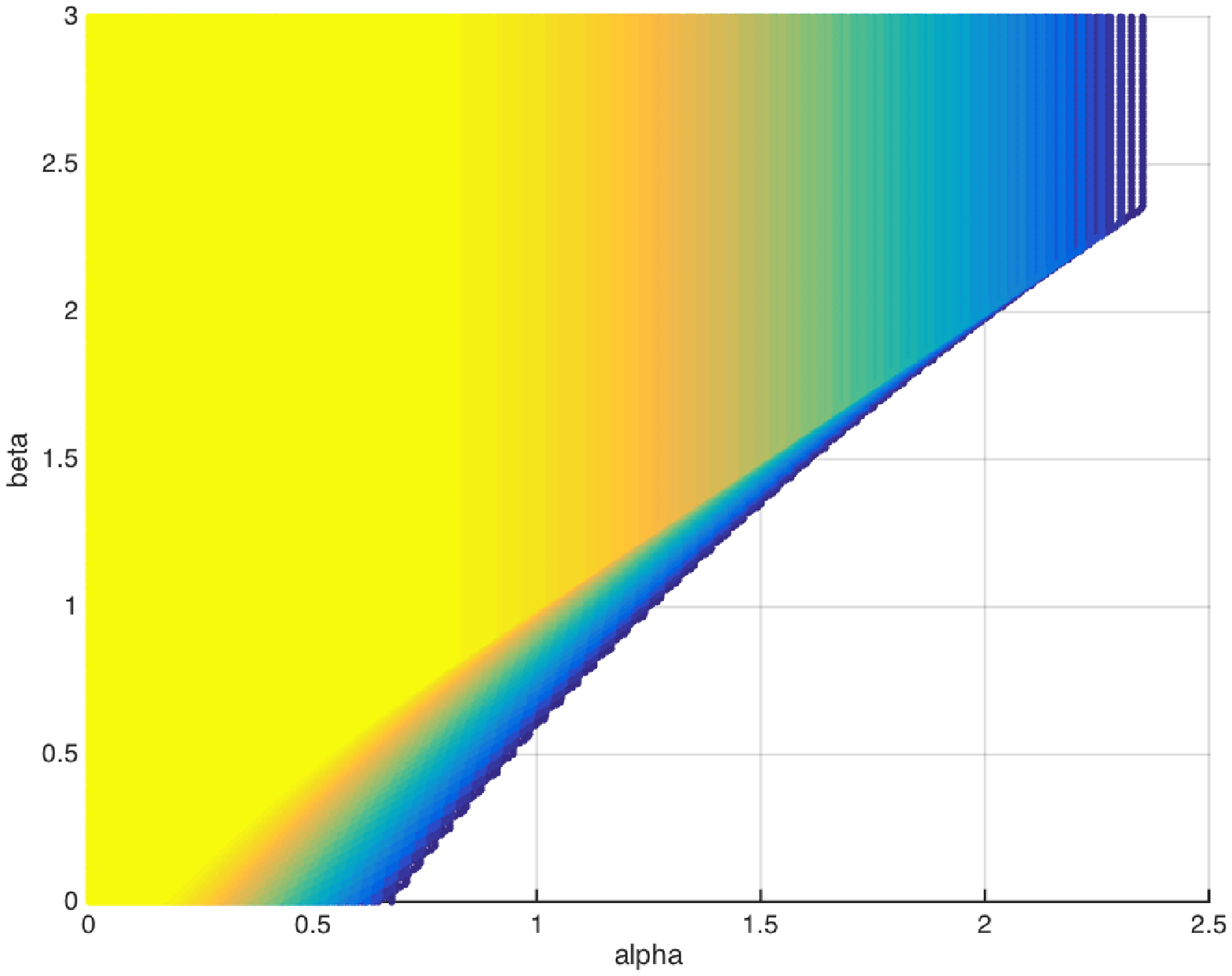}\label{fig:BSC random access alpha beta}}
\caption{Capacity region of slotted bursty and strongly asynchronous BSC with random access with cross over probability $\delta =0.11$.}
 \end{figure*}
 \end{example}
\section{Conclusion}\label{sec:conclusion}
In this paper we study a slotted bursty and strongly asynchronous discrete memoryless channel  where a user transmits a randomly selected message among $M_n=e^{nR}$ messages in each one of the $K_n=e^{n\nu}$ randomly selected blocks of the available $A_n=e^{n\alpha}$ blocks. We  derive the upper and lower bounds on the  trade-off among $(R, \alpha, \nu)$ by finding  achievability and converse bounds where we analyze an optimal Maximum Likelihood decoder in the converse.  For the case that the number of transmissions of the user is not fixed and the user may access the channel with probability $e^{-n\beta}$, we again provide upper and lower bounds on the trade-off between $(R, \alpha, \beta)$.

\section{Acknowledgement} 
The work of the authors was partially funded by NSF under award 1422511. The contents of this article are solely the responsibility of the authors and do not necessarily represent the official views of the NSF.

\appendix

\subsection{Proof of~\eqref{eq:first lower bound}}
\label{app:prob union less than T}
The main trick in the proof of~\eqref{eq:first lower bound} is to find an equivalent event and lower bound the probability of  that event instead. In this regard we have
\begin{align}
&
\prob \left[\bigcup_{i\in [K_n]}\frac{1}{n}\log \frac{Q(Y_i^n|x^n_i(m_i))}{Q_{\star^n}(Y_i^n)}\leq T 
{  } \right]\label{eq:original equiv}
\\&
= \prob[{Z}_1\geq 1]
\label{eq:equiv prob}
\\
&\geq 1-\frac{\text{Var}[Z_1]}{\mathbb{E}^2[Z_1]}=1-\frac{\sum_{i=1}^{K_n}p_i(1-p_i)}{\left( \sum_{i=1}^{K_n}P_i\right)^2}\geq 1-\frac{1}{\sum_{i=1}^{K_n}p_i}
\label{eq:second moment}
\\
&\geq 1-e^{-n\left(\nu -D\left(Q_{\lambda_i^\star} \parallel Q|P_{i^\star} \right) \right)},
\label{eq:choice i star}
\end{align}
where we define
\begin{align}
Z_1&:= \sum_{i=1}^{K_n}  {\xi}_i,\qquad  {\xi}_i\sim \text{Bernoulli}(p_i),
\notag\\
p_i&:=Q_{x_i^n(m_i)}\left[\frac{1}{n}\log \frac{Q(Y_i^n|x^n_i(m_i))}{Q_{\star^n}(Y_i^n)}\leq T  \right]\notag\\
p_i&\geq Q_{x_i^n(m_i)}\left[Y_i^n\in T_{Q_{\lambda_i}}(x_i^n(m_i)) \right]=e^{-nD(Q_{\lambda_i}\parallel Q\vert P_i)}.\label{eq:special T}
\end{align}
The equality in~\eqref{eq:equiv prob} is due to the equivalence of the events to the ones in~\eqref{eq:original equiv}
and ~\eqref{eq:second moment} is by Chebyshev's inequlity. The inequality in~\eqref{eq:choice i star} is by the choice of $i^\star$ in~\eqref{eq:i^star} and finally~\eqref{eq:special T} is true because of the special choice of $T=D(Q_{\lambda_{i }}\parallel Q_\star|P_{i })-D(Q_{\lambda_{i }}\parallel Q|P_{i })$.
\subsection{Proof of~\eqref{eq:second lower bound}}
\label{app:union prob messages}

To find a lower bound on the term in~\eqref{eq:err lb 2}, we proceed as before by writing 
\begin{align}
 &\prob \left[\bigcup_{j\in[K_n+1:A_n]} \bigcup_{\substack{m\in [M_n]}} \frac{1}{n}\log\frac{Q\left(Y_j^n|x_{j}^n(m)\right)}{Q_{\star^n}(Y_j^n)}\geq T {  } \right]
 \\&= 
 \prob\left[ Z_2 \geq 1\right]
 \label{eq:equiv prob z2}
 \\&
 \geq 1-\frac{\text{Var}[Z_2]}{\mathbb{E}^2[Z_2]}
 \label{eq:second moment z2}=1-\frac{\sum_{j=K_n+1}^{A_n}q_j(1-q_j)}{\left( \sum_{j=K_n+1}^{A_n}q_j\right)^2}\geq 1-\frac{1}{\sum_{j=K_n+1}^{A_n}q_j}
 \\&
\geq 1-\exp\left\{-n\left(\alpha+R\mathbbm{1}_{\left\{R<I(P, Q_{\lambda_{i^*}}) \right\}}-D(Q_{\lambda_{i^*}}\parallel Q_\star|P_{i^*})\right)\right\} 
,\notag
\end{align}
where we have defined
\begin{align}
 Z_2&:=\sum_{\substack{j\in[K_n+1:A_n]}} \zeta_{j},\qquad
 \zeta_{j}\sim \text{Bernoulli}(q_{j}),
\notag\\
 q_{j}&:=Q_{\star^n}
 \left[\bigcup_{\substack{m\in [M_n]}} \frac{1}{n}\log \frac{Q(Y_j^n|x_{j}^n(m))}{Q_{\star^n}(Y_j^n)}\geq T 
 {  } \right],
\label{eq:lb false alarm}\\
 & q_j \geq \exp\left\{n\left(R\mathbbm{1}_{\left\{R<I(P, Q_{\lambda_j}) \right\}}-D(Q_{\lambda_{j}}\parallel Q_\star|P_{j})\right)\right\}\label{eq:zeta prob}.
\end{align}
The equality in~\eqref{eq:equiv prob z2} is true because the two events in the probabilities are the same and the first inequality in~\eqref{eq:second moment z2} is by the Chebyshev inequality. 
The inequality in~\eqref{eq:zeta prob} is proved in Appendix~\ref{app:lower bound union messages}. 
We should  note that $\zeta_{j}, j \in[K_n+1:A_n],$ 
are independent since $Y_j^n, j \in[K_n+1:A_n]$ are independent. 

\subsection{Lower bound 
in~\eqref{eq:zeta prob}.}\label{app:lower bound union messages} 

We first define a new typical set $T_{\Qle^{n}}^{\delta}$ as follows.
\begin{definition}
For $\epsilon$ and $\delta$ define
\begin{align*}
&T_{\Qle^{n}}^\delta(x^n):=\left\{
y^n:\sum_{a,b}\frac{1}{n}\mathcal{N}\left(a,b\vert x^n,y^n\right) \log \frac{Q(b|a)}{Q_{\star}(b)}\geq T,
\right. \\
 &\left.\qquad
\left|\frac{1}{n}\mathcal{N}\left(a,b\vert x^n,y^n \right)-P(a)Q_{\lambda+\epsilon}(b|a)\right|< \delta,  \forall (a,b) \in \mathcal{X} \times \mathcal{Y}
\right\}.
\end{align*}
\end{definition}
The new constraint 
\[\sum_{a,b}\frac{1}{n}\mathcal{N}\left(a,b\vert x^n,y^n\right) \log \frac{Q(b|a)}{Q_\star(b)}\geq T\]
 that we included in the typical set definition ensures that all the sequences $y^n$ that belong to $T_{\Qle^{n}}^{\delta}$ will also satisfy 
 \[\frac{1}{n}\log \frac{Q(y^n|x^n)}{Q_{\star^n}(y^n)}\geq T.\]
In addition, define
\[\Delta:= \sum_{a,b}P(a)Q_{\lambda+\epsilon}(b|a) \log \frac{Q(b|a)}{Q_\star(b)}-T,
\]
where $\Delta>0$ since 
\[T=\sum_{a,b}P(a)Q_{\lambda}(b|a) \log \frac{Q(b|a)}{Q_\star(b)} \]
  is decreasing in $\lambda$~\cite{blahut}. 
By the Law of Large Numbers
\begin{align*}
\Qle^{n}\left[\left|\frac{1}{n}\mathcal{N}\left(a,b\vert x^n,Y^n\right)-P(a)\Qle(b|a) \right|>\delta |x^n\right]\to 0
\end{align*}
and 
\begin{align*}
\Qle^{n}\left[\sum_{a,b}\frac{1}{n}\mathcal{N}\left(a,b\vert x^n,Y^n\right)\log \frac{Q(b|a)}{Q_\star(b)}\geq T |x^n \right]\to 0 \end{align*}
and hence for any $\delta_1>0$ there exists $n_1$ such that for all $n\geq n_1$ we have
\begin{align}
\Qle^{n}\left[T_{Q_{\lambda+\epsilon}}^\delta(x^n)|x^n \right]>1-\delta_1.\label{eq:T_Qle}
\end{align}
Moreover, assume that $D_{Q^n_{\lambda+\epsilon}}(m)$ is the optimal (and disjoint) decoding region for message $m$, whose codeword is passed through the channel $Q^n_{\lambda+\epsilon}$. We also denote the average probability of decoding error associated with channel $Q^n_{\lambda+\epsilon}$ to be 
\[P_e^{(n)}(Q_{\lambda+\epsilon}) := \frac{1}{e^{nR}}\sum_{m=1}^{e^{nR}}\sum_{y^n\in D^c_{Q^n_{\lambda+\epsilon}}(m)}Q^n_{\lambda+\epsilon}(y^n).\]
Now, if we drop half of the codewords in $\left( x^n(1),\ldots, x^n(M_n)\right)$ with the largest probability of the error, the remaining half must must all satisfy 
\begin{align}
Q_{\lambda+\epsilon}^{n} \left[ {  Y}^n \not \in D_{Q_{\lambda+\epsilon}^{n}}(m)|x^n(m) \right]<2P_e^{(n)}(Q_{\lambda+\epsilon})
;\label{eq:prob not in dec region v2}
\end{align}
otherwise, the average probability of error for the decoding regions $D_{Q_{\lambda+\epsilon}^{n}}(m)$ will be larger than $P_e^{(n)}(Q_{\lambda+\epsilon})$ and we reach a contradiction. Henceforth we restrict our attention to this half of the codebook (which without loss of generality we assume is the first $\frac{M_n}{2}$ codewords).

As the result for the optimal decoding regions $D_{\Qle^n}(m)$ of channel  channel $\Qle^n$ and by~\eqref{eq:T_Qle} and~\eqref{eq:prob not in dec region v2} we have
\begin{align}
\Qle^{n}\left[T_{\Qle^{n}}^\delta\left(x^n(m)\right) \cap D_{\Qle^{n}}(m)|x^n(m)\right]\geq 1-\delta_1-2P_e^{(n)}(Q_{\lambda+\epsilon}).\label{eq: e delta}
\end{align}
In addition, we can conclude from~\cite[Lemma 10]{polyanskiy} that for any two distributions $P_1^n, P_2^n$ and any event $A$ such that
\[P_1^n(A)\geq \alpha,\]
we have
\begin{align}
P_2^n(A) \geq \beta_{\alpha}(P_1^n, P_2^n)\geq \frac{\alpha}{2}\exp \left\{-nD(P_1\parallel  P_2)
\right\}.\label{eq:polyanskiy conv}
\end{align}
In case the lower bound given in~\eqref{eq: e delta}, i.e. $1-\delta_1-P_e^{(n)}(Q_{\lambda+\epsilon})$, is positive (which we discuss shortly) and by~\eqref{eq:polyanskiy conv} we can write
\begin{align}
&Q_{\star^n}\left[\bigcup_{\substack{m\in [{M_n}]}} \frac{1}{n}\log \frac{Q(Y_i^n|x^n(m))}{Q_{\star}^{n}(Y_i^n)}\geq T \right]
\geq Q_{\star^n}\left[\bigcup_{\substack{m\in [\frac{M_n}{2}]}} T_{\Qle^{n}}^\delta\left(x^n(m)\right) \right]\notag\\
&\geq Q_{\star^n}\left[\bigcup_{\substack{m\in [\frac{M_n}{2}]}} T_{\Qle^{n}}^\delta\left(x^n(m)\right) \cap D_{\Qle^{n}}(m)\right]\notag\\
&=\sum_{m=1}^{\frac{M_n}{2}} Q_{\star^n}\left[T_{\Qle^{n}}^\delta\left(x^n(m)\right) \cap D_{\Qle^{n}}(m)\right]\notag\\
&\geq \sum_{m=1}^{\frac{M_n}{2}} \frac{1-\delta_1-2P_e^{(n)}(Q_{\lambda+\epsilon})}{2} e^{-nD(\Qle \parallel Q_\star|P)
}\notag\\
&\doteq e^{nR}e^{-nD(\Qle \parallel Q_\star|P)}.\label{eq:final lb}
\end{align}
In addition, due to continuity of the divergence, as $\epsilon\to 0$, we have 
\[D(\Qle \parallel Q_\star|P) \to D(Q_\lambda\parallel Q_\star|P).\]
We now discuss the case that $1-\delta_1-P_e^{(n)}(Q_{\lambda+\epsilon})$ is positive. A sufficient condition for $1-\delta_1-P_e^{(n)}(Q_{\lambda+\epsilon})$ to be positive is that $P_e^{(n)}(Q_{\lambda+\epsilon})$ vanishes as $n\to \infty$. This is true if 
\[R<I(P,Q_{\lambda +\epsilon}).\]
If, on the other hand $R\geq I(P,Q_{\lambda+\epsilon})$, we still can lower bound~\eqref{eq:lb false alarm}  by
\begin{align*}
Q_{\star^n}\left[\bigcup_{m\in [M_n]}\frac{1}{n} \log \frac{Q(Y_j^n|x^n(m))}{Q_{\star^n}(Y_j^n)}\geq T \right] &\geq Q_{\star^n}\left[\frac{1}{n} \log \frac{Q(Y_j^n|x^n(1))}{Q_{\star^n}(Y_j^n)}\geq T \right] \\
&\geq Q_{\star^n}\left[Y_j^n \in T_{Q_\lambda}^\delta(x^n(1)) \right] \\
&\geq e^{-nD(Q_{\lambda}\parallel Q_\star\vert P)}.
\end{align*}

\subsection{Proof of Lemma~\ref{lemma:lower envelope}} \label{app:region}
We provide the proof for a binary alphabet $\mathcal{X}=\{a,b\}$ in a proof by contradiction. 
The proof for the general $\vert \mathcal{X}\vert>2$ is a straightforward generalization. 
For $x=a,b$ define 
\begin{align*}
E_0^{(x)}(\lambda_x)&:=D\left(Q_{\lambda_x}\parallel Q_\star \right), \\
E_1^{(x)}(\lambda_x)&:=D\left(Q_{\lambda_x}\parallel Q_x \right).
\end{align*}
Assume that the claim of the Lemma~\ref{lemma:lower envelope} is not valid and hence there exists $(\lambda_a,\lambda_b,{\widetilde{\lambda}} )\in[0,1]^3$ such that 
\begin{align*}
D(Q_{\lambda_x}\parallel Q| P)<D(Q_{\widetilde{\lambda}}\parallel Q |P),\\
D(Q_{\lambda_x}\parallel Q_\star |P)<D(Q_{\widetilde{\lambda}}\parallel Q_\star |P),
\end{align*}
or equivalently
\begin{subequations}
\begin{align}
&
\rho E_1^{(a)}(\lambda_a)+\bar{\rho} E_1^{(b)}(\lambda_b) < 
\rho E_1^{(a)}({\widetilde{\lambda}})  +\bar{\rho} E_1^{(b)}({\widetilde{\lambda}} ),
\label{eq:rearrange 1}
\\
&
\rho E_0^{(a)}(\lambda_a)+\bar{\rho} E_0^{(b)}(\lambda_b) <
\rho E_0^{(a)}({\widetilde{\lambda}} )  +\bar{\rho} E_0^{(b)}({\widetilde{\lambda}} ),
\label{eq:rearrange 2}
\end{align}
where $\rho:= \PP(x=a)$ and $\bar{\rho}=1-\rho=\PP(x=b)$.
\label{eq:rearrange}
\end{subequations}
By~\cite[Theorem 2]{blahut} we can  exclude the cases where $\lambda_a, \lambda_b< {\widetilde{\lambda}}  $ and $\lambda_a, \lambda_b>{\widetilde{\lambda}} $ and assume $\lambda_a< {\widetilde{\lambda}}< \lambda_b$, {which implies
\begin{align*}
E_1^{(x)}(\lambda_a)>E_1^{(x)}({\widetilde{\lambda}})>E_1^{(x)}(\lambda_b),\\
E_0^{(x)}(\lambda_a)<E_0^{(x)}({\widetilde{\lambda}})<E_0^{(x)}(\lambda_b),
\end{align*}
 for $x\in\{a,b\}$.} 
Hence, by rearranging~\eqref{eq:rearrange} and by dividing the two equations, we get 
\begin{align}
\frac{\left(E_1^{(a)}(\lambda_a)-E_1^{(a)}({\widetilde{\lambda}} )\right)  }{\left(E_0^{(a)}(\lambda_a)-E_0^{(a)}({\widetilde{\lambda}} )\right)}
>
\frac{\left(E_1^{(b)}({\widetilde{\lambda}} )  -E_1^{(b)}(\lambda_b)\right)}{\left(E_0^{(b)}({\widetilde{\lambda}} )  -E_0^{(b)}(\lambda_b)\right)}.
\label{eq:slope}
\end{align}
{Note since the $\left(E_0^{(x)}(\lambda), E_1^{(x)}(\lambda)\right)$ curve is convex and strictly decreasing, we have
\begin{align}
\frac{\partial E_1^{(a)}\left(E_0^{(a)}(\lambda)\right)}{\partial \lambda}\big|_{\lambda={\widetilde{\lambda}}} \geq \frac{ \left(E_1^{(a)}(\lambda_a)-E_1^{(a)}(\lambda)\right)}{\left(E_0^{(a)}(\lambda_a)-E_0^{(a)}(\lambda)\right)},\label{eq:slope contrast 1}\\
\frac{\left(E_1^{(b)}(\lambda)-E_1^{(b)}(\lambda_b)\right)}{\left(E_0^{(b)}(\lambda)-E_0^{(b)}(\lambda_b)\right)}\geq \frac{\partial E_1^{(b)}\left(E_0^{(b)}(\lambda)\right)}{\partial \lambda}\big|_{\lambda={\widetilde{\lambda}} },\label{eq:slope contrast}
\end{align}
where $ \frac{\partial E_1^{(x)}\left(E_0^{(x)}(\lambda)\right)}{\partial \lambda}$ is the slope of the $\left(E_0^{(x)}(\lambda), E_1^{(x)}(\lambda)\right)$, which can be visually seen in Fig.~\ref{fig:l_a l_star}.
However, according to~\cite[Theorem 6]{blahut}, the slope of the $\left(E_0^{(x)}(\lambda), E_1^{(x)}(\lambda)\right)$ curve at $\lambda={\widetilde{\lambda}} $ is equal to $\frac{{\widetilde{\lambda}}-1}{{\widetilde{\lambda}} }$ and is independent of $x$. 

Putting~\eqref{eq:slope},~\eqref{eq:slope contrast 1} and~\eqref{eq:slope contrast} together, we reach a contradiction and the proof is complete.
\begin{figure}[htbp]
\centering
\includegraphics[width=.5\textwidth]{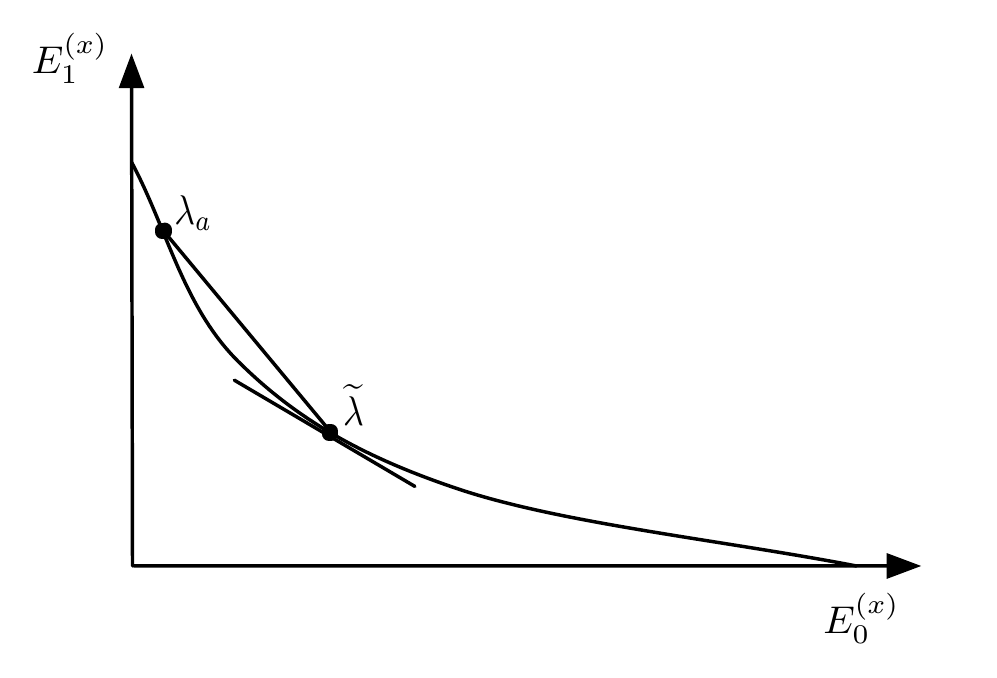}
\caption{Slope at $\lambda={\widetilde{\lambda}}$ is larger than the slope of the line between $\lambda_a$ and ${\widetilde{\lambda}}$.}
\label{fig:l_a l_star}
\end{figure}

\subsection{Proof of~\eqref{eq:conv p lb}}
\label{app:conv p lb calculation}
Note that
\begin{align*}
\sum_{k=1}^{A_n-1}\binom{A_n}{k}p^k (1-p)^{A_n-k} \frac{1}{k} &= \frac{1}{A_n+1}\sum_{k=1}^{A_n-1}\binom{A_n+1}{k+1}p^k (1-p)^{A_n-k} \frac{k+1}{k}\\
&\leq \frac{2}{A_n+1}\sum_{k=1}^{A_n-1}\binom{A_n+1}{k+1}p^k (1-p)^{A_n-k}\\
&\leq \frac{2}{p(A_n+1)}\sum_{j=0}^{A_n+1}\binom{A_n+1}{j}p^j (1-p)^{A_n+1-j}\\
&= \frac{2}{p(A_n+1)}\leq \frac{2}{p A_n},
\end{align*}
and similarly
\begin{align*}
\sum_{k=1}^{A_n-1}\binom{A_n}{k}p^k (1-p)^{A_n-k} \frac{1}{A_n-k}\leq \frac{2}{(1-p)A_n}.
\end{align*}
\bibliography{refs}
\bibliographystyle{IEEEtran}
\end{document}